\documentclass[%
 twocolumn,
 floatfix,
 superscriptaddress,
 amsmath,amssymb,
 aps,
]{revtex4-2}

\usepackage{booktabs}
\usepackage[x11names]{xcolor}
\usepackage{graphicx}
\usepackage{dcolumn}
\usepackage{bm}
\usepackage{hyperref}
\usepackage{cleveref}
\usepackage[mathlines]{lineno}
\usepackage{todonotes}
\usepackage{subcaption}
\usepackage[normalem]{ulem}
\usepackage{ragged2e}
\usepackage[x11names]{xcolor}
\usepackage{custom}

\newcommand{\MeV}{\, \rm{MeV}}

\newcommand{\coef}{\alpha}

\begin{document}


\title{Locating the QCD critical point 
through contours of constant entropy density}

\author{Hitansh Shah}
\affiliation{
 Department of Physics, University of Houston, Houston, TX 77204, USA
}

\author{Mauricio Hippert}
\affiliation{Instituto de Física, Universidade do Estado do Rio de Janeiro, Rua São Francisco Xavier, 524, Rio de Janeiro, RJ, 20550-013, Brazil}

\author{Jorge Noronha}
\affiliation{Illinois Center for Advanced Studies of the Universe \& Department of Physics,
University of Illinois at Urbana-Champaign, Urbana, IL 61801-3003, USA}

\author{Claudia Ratti}
\affiliation{
 Department of Physics, University of Houston, Houston, TX 77204, USA
}

\author{Volodymyr Vovchenko}
\affiliation{
 Department of Physics, University of Houston, Houston, TX 77204, USA
}

\date{\today}%

\begin{abstract}
We propose a new method to investigate the existence and location of the conjectured high-temperature critical point of strongly interacting matter via contours of constant entropy density. 
By approximating these lines as a power series in the baryon chemical potential $\mu_B$, one can extrapolate them from first-principle results at zero net-baryon density, and use them to locate the QCD critical point, including the associated first-order and spinodal lines. 
As a proof of principle, we employ currently available continuum-extrapolated lattice data from the Wuppertal--Budapest collaboration to find a critical point at a temperature and a baryon chemical potential of $T_c = 114.3 \pm 6.9$ MeV and $\mu_{B,c} = 602.1 \pm 62.1$ MeV, respectively, at expansion order $\mathcal{O}(\mu_B^2)$. We advocate for a more precise determination of the required expansion coefficients via lattice QCD simulations as a means of pinpointing the location of the critical endpoint in the phase diagram of strongly interacting matter. 
\end{abstract}

\maketitle


%
\paragraph*{\bf Introduction.}
One of the most challenging questions in the field of quantum chromodynamics (QCD) concerns the existence of a first-order phase transition at finite baryon chemical potential $\mu_B$ and high temperature $T$, separating the hadronic and quark-gluon plasma phases. 
From state of the art lattice QCD calculations, we know that the transition from hadronic matter to deconfined quark gluon plasma at $\mu_B = 0$ is an analytic, smooth crossover \cite{Aoki:2006br}, taking place at a pseudocritical temperature of 155 -- 158 MeV \cite{Borsanyi:2020fev, HotQCD:2018pds}. 
Because direct lattice QCD simulations are not feasible at finite density due to the fermion sign problem, whether a first-order phase transition occurs at large chemical potential remains an open question. 
Extrapolations of lattice results to small baryon densities indicate that the analytic crossover persists at least up to $\mu_B/T \leq 3$~\cite{Vovchenko:2017gkg,Borsanyi:2020fev,Bollweg:2022rps}, although a hint of a narrowing of the crossover towards large $\mu_B$ was observed for the first time in \cite{Borsanyi:2024xrx}. Several effective QCD theories predict a first-order phase transition at large $\mu_B$, originating from a critical point (CP) where the transition is of second order (for recent reviews, see \cite{Bzdak:2019pkr,Du:2024wjm}). 
While in the past the theoretical predictions for the location of the QCD critical point were scattered all over the QCD phase diagram, more recent ones seem to concentrate around a considerably narrower region, with chemical potentials in the range $400\leq\mu_B\leq650$ MeV~\cite{Fu:2019hdw,Gunkel:2021oya,Gao:2020fbl,Hippert:2023bel,Basar:2023nkp,Clarke:2024ugt}. 

The universal features and peculiar behavior expected near the QCD critical point make a compelling case for its experimental search in heavy-ion collisions,
in particular utilizing event-by-event fluctuations in multiplicities of produced particles~\cite{Stephanov:1999zu,Hatta:2003wn,Stephanov:2008qz}.
The experimental search for the QCD critical point is one of the primary goals of the Beam Energy Scan program at RHIC, which finished its runs in 2021~\cite{Luo:2017faz,Bzdak:2019pkr}.
Proton cumulants are expected to show characteristic signatures of critical behavior, although interpreting experimental results in this context is challenging.
The data from the STAR Collaboration~\cite{STAR:2020tga,STAR:2021iop}, including the new preliminary data from BESII on proton number factorial cumulants~\cite{PandavCPOD2024}, hint at an interesting behavior for energies $\sqrt{s_{NN}} \leq 20$ GeV. These features cannot be explained by non-critical baselines~\cite{Braun-Munzinger:2020jbk,Vovchenko:2021kxx}, even though no conclusive statement on the critical point existence/location has been made so far either using the proton cumulants, or other considerations such as finite-size scaling~\cite{Fraga:2011hi,Lacey:2014wqa,Sorensen:2024mry}.
A common strategy to sidestep the obstacles for a direct lattice QCD determination of the equation of state at high baryon densities is to rely on extrapolations from zero or imaginary chemical potentials, such as analytic continuation and Taylor expansion~\cite{Allton:2003vx,Allton:2005gk,Borsanyi:2012cr,Bazavov:2017dus,Bollweg:2022fqq}. 
However, these methods often fail at large densities. Especially, the Taylor expansion of thermodynamic pressure in powers of $\mu_B/T$ is restricted by non-analyticities in the complex plane, including the CP itself \cite{Gavai:2004sd}, and hence cannot reach the CP.
Recent approaches to extract information about the CP from lattice QCD results analyze the expected Lee-Yang edge singularities at complex baryon chemical potential and extrapolate down in temperature to locate the CP through their crossing of the real $\mu_B$ axis~\cite{Basar:2023nkp,Clarke:2024ugt}.

\begin{figure*}[t]
    \includegraphics[width=\linewidth]{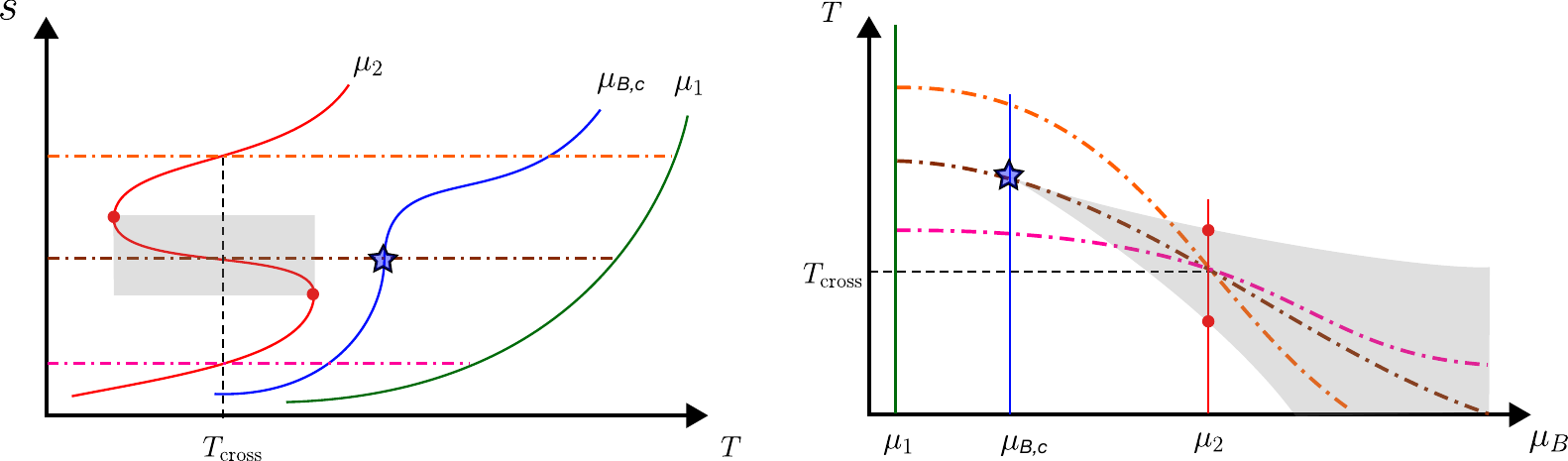}
    \caption{
    \justifying \small 
    Left: entropy as a function of the temperature, at three different chemical potentials. In this scheme, $\mu_1<\mu_{B,c}<\mu_2$. 
    Right: corresponding constant-entropy contours in the $(T,~\mu_B)$ plane. The blue star indicates the critical point, while the shaded area corresponds to the spinodal region. The red dots indicate the spinodal points for $\mu_B=\mu_2$.  
    }
    \label{fig:cartoon_firstorder}
\end{figure*}
In this Letter, we present a new method to 
locate the CP in the QCD phase diagram
which utilizes first-principle lattice QCD input,
exploiting the fact that the entropy density $s$ is a multi-valued function in the region of a first-order phase transition, 
reflecting phase coexistence. 
The existence of a CP thus implies crossings of constant entropy density contours.
Since lattice QCD simulations can only be performed at vanishing baryon chemical potentials, we extrapolate the entropy density contours from $\mu_B = 0$ by approximating them with an expansion in powers of $\mu_B$.
In contrast to other expansions used in the literature, this scheme explicitly permits the crossings of these contours.

As we discuss in the following, it describes the CP and associated mixed-phase region in a mean-field approximation.
Thus, the method allows for the determination of whether and where a CP is found in the phase diagram, as well as yielding spinodal and coexistence lines for the QCD phase transition. 
To utilize the first-principles input, we extract expansion coefficients for the lines of constant entropy density from state-of-the-art lattice QCD results from the Wuppertal--Budapest collaboration.
Our reconstructed entropy density agrees with the lattice QCD results from \cite{Borsanyi:2021sxv} up to $\mu_B/T \leq 3.5$. 
At larger chemical potentials, it develops the S-shape (Maxwell loops) of a first-order phase transition, indicating the emergence of the CP singularity.
The results remain consistent with the HotQCD Collaboration’s input for entropy density \cite{HotQCD:2014kol}.
As a consistency check, we find that the method successfully locates the known CP in the holographic model developed in \cite{Critelli:2017oub}.
Our method also shows indications for a CP at imaginary $\mu_B$ close to where the endpoint of the Roberge-Weiss transition is expected.

\medskip
\paragraph*{\bf  Crossings of entropy contours.}\label{sec:intuition}

The intuitive picture behind our approach is grounded in the multi-valued behavior of the entropy density $s$ around a first-order phase transition. 
Strictly speaking, entropy density is a single-valued function of $T$ and $\mu_B$ in equilibrium, with the only exception being the phase coexistence line in the thermodynamic limit.
However, the multi-valuedness can appear by considering metastable states.
The most prominent example is the van der Waals equation of state (and other analytic models), which neglects fluctuations of the order parameter and describes the first-order phase transition in the mean field approximation.
The Maxwell construction is then employed to eliminate the multi-valuedness and determine the equilibrium state. Our method here is based on analytic extrapolation of constant entropy contours from crossover toward a CP and yields a similar structure, which we discuss in detail below.

The expected relationship between the temperature and entropy density in the presence of the QCD CP at finite $\mu_B$, in the mean-field approximation,
is illustrated in Fig.~\ref{fig:cartoon_firstorder}. 
The curves on the left panel show $s$ \textit{versus} $T$ for different values of the chemical potential: the critical value $\mu_{B,c}$,  $\mu_1 <\mu_{B,c}$ and  $\mu_2 > \mu_{B,c}$. 
For $\mu_B < \mu_{B,c}$, the entropy is a continuous function of the temperature, with a steep rise around the crossover region. As $\mu_B$ increases, this steep rise becomes more pronounced, making the lines of constant entropy density closer in temperature, until a diverging slope is achieved at the critical point. 
For $\mu_B> \mu_{B,c}$, the entropy can have three values for the same $(T,~\mu_B)$ pair.  Thus, up to three lines of constant $s$ can cross at the same point in the $(T,~\mu_B)$ plane.
This is shown by the horizontal dash-dotted lines depicting three different contours of entropy density.
These contours are shown in the $T$-$\mu_B$ plane in the right panel of Fig.~\ref{fig:cartoon_firstorder}, where they cross at the same point. 
In fact, crossings between contours of constant $s$ will spread through the region in the $(T,~\mu_B)$ plane between the spinodal curves. 

Precisely at the spinodal points, the temperature as a function of $s$ displays a local maximum or minimum (see Fig.~\ref{fig:cartoon_firstorder}), so that      $\left(\partial T/ \partial s\right)_{\mu_B} = 0$.  
As the critical point is approached, the local maximum and minimum coalesce, leading to an inflection point with $\left(\partial^2 T/ \partial s^2\right)_{\mu_B} = 0$, while still satisfying $\left(\partial T/ \partial s\right)_{\mu_B} = 0$. 
The location of the CP is thus determined by a pair of equations
\begin{equation}
\label{eq:TCP}
    \left(\partial T/ \partial s\right)_{\mu_B} = 0, \qquad \left(\partial^2 T/ \partial s^2\right)_{\mu_B} = 0.
\end{equation}
Similar arguments hold for other quantities that become multivalued around the first-order phase transition and may serve as  order parameters. 
For instance, the CP of a liquid-gas transition corresponds to the vanishing first and second derivatives of the pressure with respect to the density at constant temperature.
We note that the relations~\eqref{eq:TCP} hold at the CP not only in the mean-field limit, but also in the 3D Ising universality class expected in QCD~\cite{Widom:1965pus}.

\medskip
\paragraph*{\bf New expansion.}
We describe the contours of constant entropy density in terms of the function $T_s(\mu_B;T_0)$, defined such that
\begin{equation}
\label{eq:scheme}
    s\big[T_s(\mu_B; T_0),\mu_B\big] = s\big(T_0,\mu_B=0\big)\,.
\end{equation}
That is, for increasing values of $\mu_B$, one can keep the entropy fixed at $s_0 = s(T_0,\mu_B=0)$ by choosing the temperature $T = T_s(\mu_B; T_0)$.
Anchoring the entropy density contour at $\mu_B = 0$ will allow us to utilize the lattice QCD input.
The expansion of $T_s(\mu_B; T_0)$ in powers of $\mu_B$ reads
\begin{equation}
    T_s(\mu_B; T_0) \approx T_0 + \sum_{n=1}^N \coef_{2n}(T_0) \,\frac{\mu_B^{2n}}{(2\,n)!} + \mathcal{O}\left(\mu_B^{2(N+1)}\right),
    \label{eq:expS}
\end{equation}
where we truncate the series at order $\mathcal{O}\left(\mu_B^{2N}\right)$. 
Of course, $T_s(\mu_B = 0; T_0) = T_0$ by definition.
Note that the expansion~\eqref{eq:expS} explicitly permits the existence of the crossings even at leading order~($N = 1$), namely $T_s(\mu_B;T_{0,1}) = T_s(\mu_B;T_{0,2})$ is possible for two different values of $T_0$.

From the function $ T_s(\mu_B; T_0)$, we can evaluate the conditions for the presence of a critical point and the shape of spinodal lines, following the discussion above. 
At the spinodal lines,
\begin{equation}
\label{eq:spinodalcond}
    \displaystyle \left(
    \frac{\partial T_s}{\partial s}\right)_{\mu_B}
    %
    = 0\,
    \quad \Rightarrow \quad
    \left(\frac{\partial T_s}{\partial T_0}\right)_{\mu_B} =0\,,
\end{equation}
where the second equation follows from the first one due to the chain rule.
At the critical point, Eq.~\eqref{eq:spinodalcond} is complemented by the condition
\begin{equation}
    \label{eq:criticalcond}
    \displaystyle \left(\frac{\partial^2 T_s}{\partial s^2}\right)_{\mu_B}
    = 0\,
    \quad \Rightarrow \quad
    \left(\frac{\partial^2 T_s}{\partial T_0^2}\right)_{\mu_B}=0\,.
\end{equation}
The new expansion is different from the commonly used Taylor expansion method. 
The Taylor method expands the thermodynamic observable of interest, typically the pressure, in powers of $\mu_B$ directly.
A truncated Taylor expansion does not permit the crossings of lines of constant observable and cannot satisfy the condition \eqref{eq:TCP} for the CP.
Our new expansion does not expand the observable directly, but rather provides a scheme to reconstruct the observable (here, the entropy density) via Eq.~\eqref{eq:scheme}.
As we show below, it can satisfy Eq.~\eqref{eq:TCP} even at leading $\mathcal{O}(\mu_B^2)$ expansion order.

\medskip
\paragraph*{\bf Expansion coefficient $\coef_{2}$.}

The expansion coefficients $\coef_{2n}$ are evaluated at $\mu_B = 0$ and can be expressed in terms of other thermodynamic quantities (see the Supplemental Material for technical details).
The expression for $\coef_{2}$ reads~
\begin{equation}
\label{eq:C2}
    \coef_{2}(T_0) = -\frac{2T_0 \chi_2^B(T_0) + T_0^2 \chi_2^{B'}(T_0)}{s'(T_0)}.
\end{equation}
Here 
$\chi_2^B = \left[ \frac{\partial^2 (p/T)^4}{\partial (\mu_B/T)^2} \right]_T$
is  the baryon number susceptibility, $s(T_0)$ is the entropy density, and $'$ indicates the temperature derivative at a constant $\mu_B$. 
In Eq.~\eqref{eq:C2}, we can see that the expansion coefficient contains not only the second derivative of pressure with respect to $\mu_B$, but also the second derivative with respect to $T$ and a mixed derivative with respect to $T$ and (twice) $\mu_B$. This reflects the fact that the present expansion is not taken at constant temperature, but rather on contours of constant entropy density---which curve towards lower temperatures at higher values of $\mu_B$---thus covering an extended region of the phase diagram. 
High-order coefficients $\coef_{2n}$ involve high-order baryon number susceptibilities and high-order temperature derivatives.
We restrict the present analysis to  $\coef_{2}$ %
and leave the inclusion of higher order terms for the future when more precise lattice results become available. 

\paragraph*{\bf Lattice QCD input.}

As follows from Eq.~\eqref{eq:C2}, the input quantities consist of the temperature dependencies of the entropy density $s(T)$ and baryon number susceptibility $\chi_2^B(T)$ at $\mu_B = 0$.
Continuum extrapolated results from the Wuppertal-Budapest collaboration have been presented in Ref.~\cite{Borsanyi:2013bia} for the former and in Ref.~\cite{Borsanyi:2021sxv} for the latter.
As our analysis relies on temperature derivatives of these two quantities, we parametrize the lattice input, based on Ref.~\cite{Kahangirwe:2024cny}.
%
We extract the values of the parameters and their full covariance matrix from the lattice results, considering strong correlations between lattice data errors of neighboring temperatures, and use this covariance matrix for Monte Carlo sampling to perform error analysis.
Additionally, we used smoothing splines to describe $s(T)$ and $\chi_2^B(T)$ and their temperature derivatives, as a more agnostic approach to employing the lattice input.
The obtained results are consistent within errors to those obtained using the parametrization. 

\paragraph*{\bf Entropy density at finite baryochemical potential.}

\begin{figure*}
    \centering
    \includegraphics[width=0.45\textwidth]{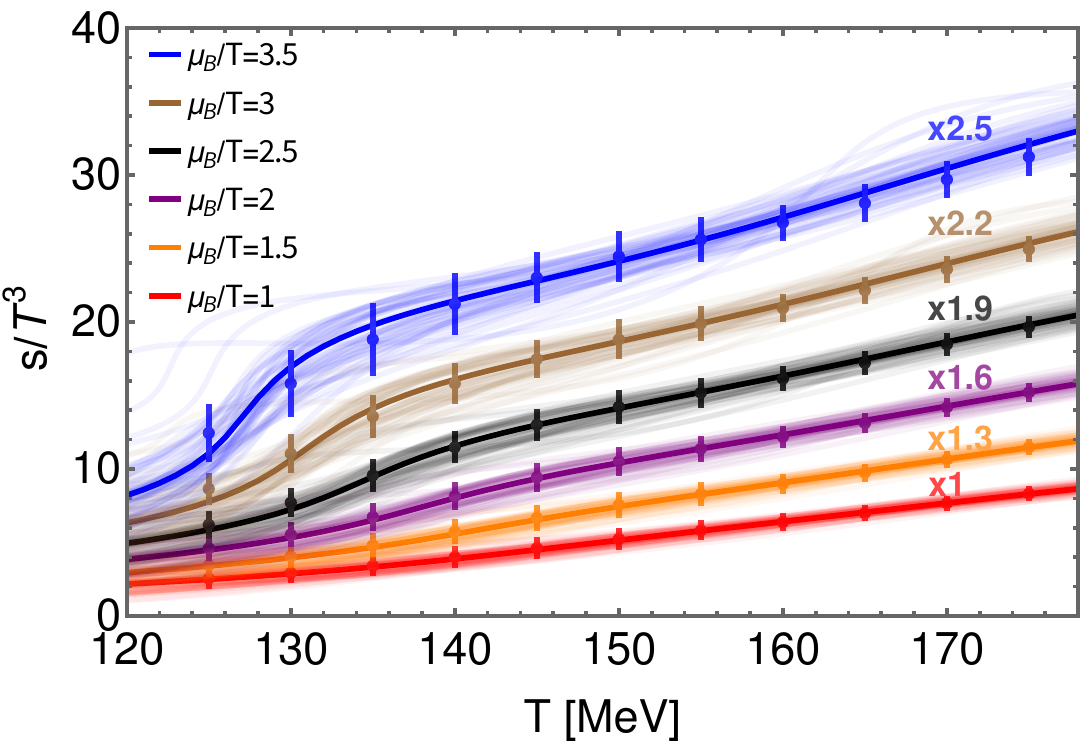}
    \hspace{1cm} 
    \includegraphics[width=0.45\textwidth]{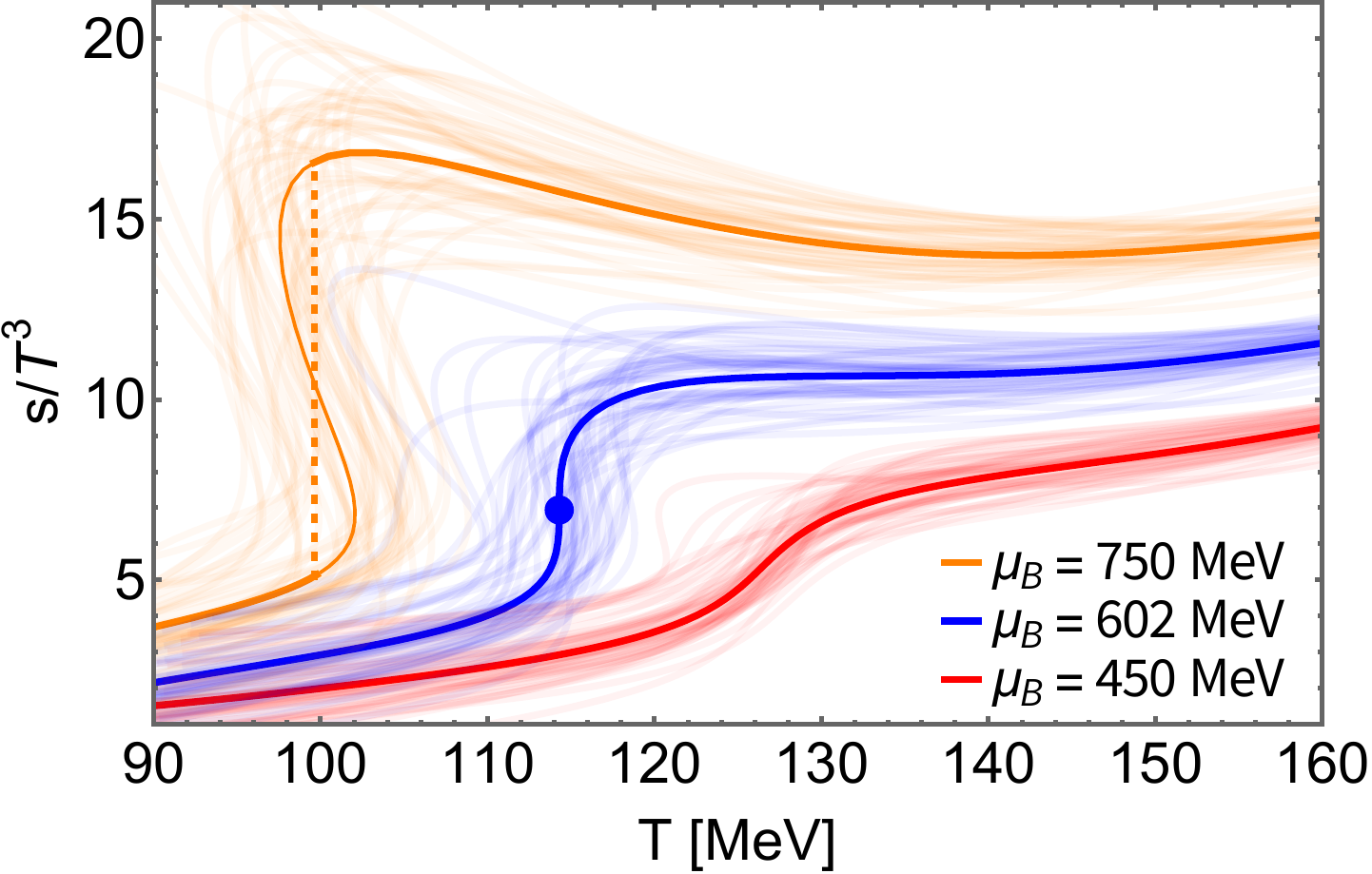}
    \caption{\justifying \small 
    Left panel: Scaled entropy density as a function of the temperature for different slices of constant $\mu_B/T$. The curves are multiplied by different factors to avoid plotting overlap. Lattice QCD data points with error bars from Ref.~\cite{Borsanyi:2021sxv} are also shown by the dots and vertical lines.
    Right panel: Scaled entropy density as a function of the temperature for different slices of constant baryochemical potential, $\mu_B = 450$~MeV~(red lines, crossover regime), $\mu_B = 602$~MeV~(blue lines, critical regime), $\mu_B = 750$~MeV~(orange lines, mixed phase regime).
    The blue circle shows the critical point.
    In both panels, the solid curves correspond to the mean parametrization of lattice QCD results, while the translucent curves reflect the error propagation of the lattice QCD input via Monte Carlo sampling.
    }
    \label{fig:sT3}
\end{figure*}

The left panel of Fig.~\ref{fig:sT3} depicts the temperature dependence of the scaled entropy density $s/T^3$ at different values of $\mu_B/T$: 1~(red), 1.5~(orange), 2~(purple), 2.5~(black), 3~(brown), and 3.5~(blue).
The results are obtained within the new expansion described above at order $\mathcal{O}(\mu_B^2)$. 
As an important cross-check, we compare our results with state-of-the-art lattice QCD data from Ref.~\cite{Borsanyi:2021sxv}.
One can see excellent agreement of our results with lattice QCD.
In addition, the hazy lines show the different Monte Carlo samples of our parametrization, reflecting the propagation of the errors from the lattice QCD input. 
The resulting spread shows good agreement with the error bars of the lattice data of Ref.~\cite{Borsanyi:2021sxv}.

The right panel of Fig.~\ref{fig:sT3} shows the behavior of $s/T^3$ at larger chemical potentials, $\mu_B = 450$~MeV~(red), 602~MeV~(blue), and 750~MeV~(orange).
One can see the development of a distinct S-shaped, multi-valued behavior of the entropy density as $\mu_B$ is increased to $\mu_B \gtrapprox 600$~MeV -- a hallmark feature of a first-order phase transition.

\paragraph*{\bf Critical point location.}

The CP is determined by a pair of equations $\left(\frac{\partial T_s}{\partial T_0}\right)_{\mu_B} = 0$ and $\left(\frac{\partial^2 T_s}{\partial T_0^2}\right)_{\mu_B} = 0$.
At order $\mathcal{O}(\mu_B^2)$, the function $T_s(\mu_B;T_0)$ reads
\begin{equation}
\label{eq:TsO2}
T_s(\mu_B;T_0) = T_0 + \coef_{2}(T_0) \frac{\mu_B^2}{2}.
\end{equation}
Denoting $\mu_{B,c}$ and $T_{0,c}$ as the values of $\mu_B$ and $T_0$ corresponding to the CP, the first equation, $\left(\frac{\partial T_s}{\partial T_0}\right)_{\mu_B} = 0$, yields the relationship between $\mu_{B,c}$ and $T_{0,c}$:
\begin{equation}
\label{eq:spinodal}
1 + \coef_2'(T_{0,c}) \frac{\mu_{B,c}^2}{2} = 0 \quad \Rightarrow \quad \mu_{B,c} = \sqrt{-\frac{2}{\coef_2'(T_{0,c})}}. 
\end{equation}
This equation also determines the spinodal lines at $\mu_B > \mu_{B,c}$ where it has two solutions with respect to $T_0$.

The second equation, $\left(\frac{\partial^2 T_s}{\partial T_0^2}\right)_{\mu_B} = 0$, corresponds to
\begin{equation}
\label{eq:T0c}
\coef_2''(T_{0,c}) = 0,
\end{equation}
which determines $T_{0,c}$.
The determination of the CP location thus proceeds by solving Eq.~\eqref{eq:T0c} for $T_{0,c}$, plugging the result into Eq.~\eqref{eq:spinodal} to determine $\mu_{B,c}$, and  computing $T_c = T_s(\mu_{B,c};T_{0,c})$ through Eq.~\eqref{eq:TsO2}. At $\mu_B > \mu_{B,c}$, the Maxwell construction of equal areas for entropy density can be utilized to determine the phase coexistence line.

Utilizing the above equations and a parametrized lattice QCD input, we obtain 
$T_{0,c} = 140.9 \pm 2.0$~MeV, and
\begin{equation}
(T_c,\mu_{B,c}) = (114.3 \pm 6.9, 602.1 \pm 62.1)~\text{MeV}
\end{equation}
for the CP location in the $T$-$\mu_B$ plane. 
The uncertainties in $T_c$ and $\mu_{B,c}$ are (anti-)correlated, with
a Pearson correlation coefficient of -0.91.
The uncertainties correspond exclusively to the Gaussian error propagation of the lattice QCD input and do not incorporate the truncation error due to terminating the expansion~\eqref{eq:TsO2} at $\mathcal{O}(\mu_B^2)$.
Furthermore, our expansion yields a CP in the mean-field universality class, as opposed to the 3D Ising universality class expected in QCD.
Our estimate thus contains additional systematic error due to the mean-field approximation. One can expect the size of this uncertainty to reflect the size of the critical scaling region.

\begin{figure}[!h]
    \centering
    \includegraphics[width=1.0\linewidth]{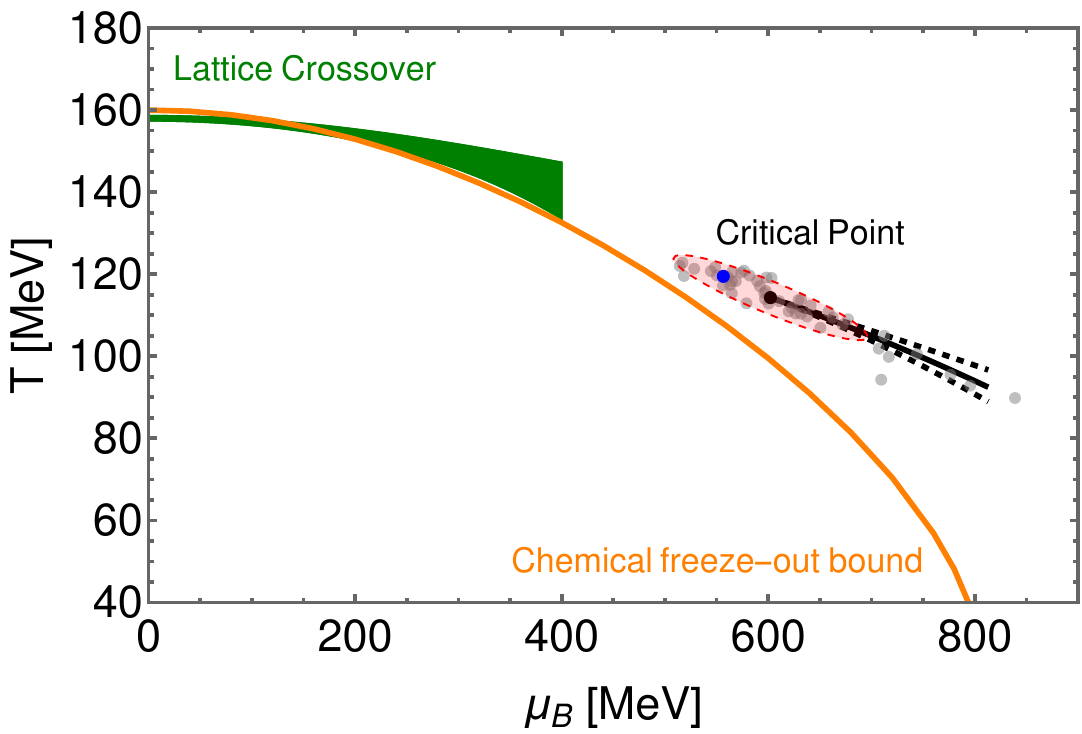}
    \caption{\justifying \small
    The solid points depict the location of the QCD critical point extracted using the contours of constant entropy density based on lattice QCD data using either parametrization~(black) or smoothing splines~(blue). 
    The dashed ellipse corresponds to the 68\% confidence interval and reflects the error propagation of the lattice input.
    The hazed gray points show the scatter of the critical points obtained by sampling the lattice QCD input.
    The solid and dashed black lines depict the coexistence curve and the spinodals, respectively.
    The green band shows the chiral crossover line from~\cite{Borsanyi:2020fev}.
    The orange line corresponds to the heavy-ion chemical freeze-out bound on the CP from~\cite{Lysenko:2024hqp}.
    }
    \label{fig:CP}
\end{figure}

The obtained CP is shown in the $T$-$\mu_B$ plane in Fig.~\ref{fig:CP}, together with the spinodals and the coexistence line.
The shown covariance ellipse corresponds to the 68\% confidence interval.
We also cross-check the error estimate by sampling the (Gaussian) log-likelihood of our lattice QCD input parametrization and depicting the resulting spread of critical points by hazy gray circles in Fig.~\ref{fig:CP}.

The obtained CP location is consistent with the possible continuation of the chiral crossover line~\cite{Borsanyi:2020fev}, shown in Fig.~\ref{fig:CP} by the green band at $\mu_B < 400$~MeV.
In the context of heavy-ion collisions, the CP should lie above the chemical freeze-out line where the hadron resonance gas model description of hadron yields applies.
The chemical freeze-out bound on the lowest CP temperature obtained in Ref.~\cite{Lysenko:2024hqp} under $\mu_Q = \mu_S = 0$ conditions appropriate for our analysis is shown in Fig.~\ref{fig:CP} by the orange line, and it lies $\approx 10$~MeV below our CP estimate in temperature.
Based on the $\mu_B$ dependence of the chemical freeze-out curve~\cite{Vovchenko:2015idt}, we find that the collision energy range of $\sqrt{s_{\rm NN}} =  4 \div 6$~GeV is in the closest vicinity to our CP location estimate.

Our results are generally compatible with other CP estimates in recent literature based on different approaches, such as the Dyson-Schwinger equation, functional renormalization group, or holography~\cite{Fu:2019hdw,Gunkel:2021oya,Gao:2020fbl,Hippert:2023bel}.
Estimates from Yang-Lee edge singularity analyses yield somewhat smaller $T$ or $\mu_B$ values~\cite{Basar:2023nkp,Clarke:2024ugt}, however they rely on high-order baryon susceptibilities that lack continuum-extrapolated data and have large systematic uncertainties and cut-off effects.
Our value of $T_c$ is considerably lower, however, than suggested by a recent finite-size scaling analysis of proton cumulants in heavy-ion collisions~\cite{Sorensen:2024mry}.

\paragraph*{\bf Consistency checks.}

Lattice QCD groups 
typically use splines to perform analyses on higher order fluctuations~\cite{Borsanyi:2023wno,Bollweg:2022rps}.
We performed a corresponding check by using a smoothed spline interpolation of the lattice data for $s(T)$ and $\chi_2^B(T)$ to calculate $\coef_2(T)$ and determine the CP~(see the Supplemental Material), which is found at $(T_c,\mu_{B,c})=(119.5 \pm 5.4,556.5 \pm 49.8) \MeV$. 
The mean value is inside the $68\%$ confidence ellipse obtained with the parametrization, as shown by the blue point in Fig.~\ref{fig:CP}.

As a different check, we utilized the parameterization of the $\mu_B = 0$ equation of state from the HotQCD Collaboration from Ref.~\cite{HotQCD:2014kol} in place of our parametrization
for the entropy density.
Using the mean values of the parameters, we obtain a CP located at $(T_c, \mu_{B,c}) = (111.4,634.8) \MeV$, 
inside our $68\%$ confidence region.

We also checked the performance of our method against solvable models. 
The holographic model~\cite{Critelli:2017oub} constrained by lattice QCD data contains a CP at $(T_c,\mu_{B,c}) \approxeq (103, 599)$~MeV~\cite{Hippert:2023bel}.
When we use the input for $s(T_0)$ and $\chi_2^B(T_0)$ to determine $\coef_2(T_0)$ in the same holographic model, we obtain a CP at $(T_c,\mu_{B,c}) \approxeq (104.5, 637.6)$~MeV
corresponding to 
a relative error of 1.4\% and 6.4\% in $T_c$ and $\mu_{B,c}$, respectively.

Another useful test is provided by the ideal HRG model, which has no CP. Thus, this model  can be used to test whether our method may produce a fake signature of a CP.
Utilizing an open source HRG code~\cite{Vovchenko:2019pjl}, we tested that the CP equation $\coef''_2(T_{0,c}) = 0$~[Eq.~\eqref{eq:T0c}] is not satisfied at any temperature $T_0$ in the range of $120 < T_0 < 200$~MeV in the ideal HRG model.
%

%

\paragraph*{\bf Roberge-Weiss endpoint.}
The equation~\eqref{eq:C2} determining the value of $T_{0}$ corresponding to a CP permits a second solution at $\tilde T_{0,c} = 160.2 \pm 1.8$~MeV.
The value of $\coef_{2}'(\tilde T_{0,c})$ at this temperature is positive and, therefore, the critical chemical potential $\tilde \mu_{B,c}$ given by Eq.~\eqref{eq:spinodal} is purely imaginary.
By calculating the critical temperature through~Eq.~\eqref{eq:TsO2} we obtain
\begin{equation}
    \tilde T_c = 197.1 \pm 7.1~\text{MeV}, \quad \frac{\tilde \mu_{B,c}}{\tilde T_c} = i (3.50 \pm 0.30).
\end{equation}
Remarkably, this value is close to where the Roberge-Weiss~(RW) transition endpoint~\cite{Roberge:1986mm,Philipsen:2014rpa,Czaban:2015sas} is expected in the plane of temperature and imaginary chemical potential, namely 
$T_{\rm RW} = 208 \pm 5$~MeV~\cite{Bonati:2016pwz,Dimopoulos:2021vrk}, and
$\mu_{B, \rm RW}/T_{\rm RW} = i \pi$~\cite{Roberge:1986mm}.
One should note that our new expansion does not explicitly incorporate the RW periodicity $\mu_B \to \mu_B + i 2 \pi T$ of the partition function and the exact relation between the extracted CP at imaginary $\mu_B$ and the RW endpoint requires further investigations.

\paragraph*{\bf Conclusion.} 
In this Letter, a novel method for determining the QCD critical point is proposed, based on the crossings of constant entropy density contours.
The expansion~\eqref{eq:expS} of the $s = \rm const.$ line away from $\mu_B = 0$ is utilized,
which overcomes the shortcomings of other expansion schemes by permitting the description of a CP and a first-order phase transition already at the leading expansion order.
Utilizing this novel expansion 
at order $\mathcal{O}(\mu_B^2)$
and the
continuum-extrapolated lattice QCD data from the Wuppertal-Budapest collaboration for entropy density $s(T)$ and baryon susceptibility $\chi_2^B(T)$ as input, we find evidence for the presence of the QCD critical point in the QCD phase diagram at $(T_c,\mu_{B,c}) = (114.3 \pm 6.9, 602.1 \pm 62.1)~\text{MeV}$, with the quoted uncertainty corresponding exclusively to the propagation of lattice QCD errors for second order baryon susceptibility and entropy density.
It remains to be seen whether the CP estimate obtained here at $\mathcal{O}(\mu_B^2)$ order reflects the true QCD CP or a breakdown of the expansion for unrelated reasons.

The analysis can be improved in the future by utilizing more precise lattice input.
The CP is determined by the behavior of the coefficient $\coef_2$ and its temperature derivatives, which are expressed in terms of entropy density and baryon susceptibility.
Therefore, a more direct determination of $\coef_2$ and its temperature derivatives on the lattice can significantly improve the statistical error.
Analysis of the truncation error due to using the expansion~\eqref{eq:expS} at order $\mathcal{O}(\mu_B^2)$ is also important, and how it may modify the result will depend on the convergence properties of the expansion.
Such an analysis will require the accurate determination of higher-order susceptibilities such as $\chi_4^B$, and high-order temperature derivatives, which are presently not all available in the continuum limit from lattice QCD.
Alternatively, one could utilize direct extrapolations of entropy density contours from zero and imaginary $\mu_B$ toward positive, real $\mu_B$, and analyze the possible emergence of their crossings.
Given that the extracted values of $T_{0,c}$ are of order $T_{0,c} \approxeq 140$~MeV, and the analysis relies on temperature derivatives, it is suggestive to extend lattice QCD simulations to lower temperatures, down to $T \approx 120$~MeV.
\paragraph*{\bf Data Availability.} The data that support the findings of
this article are openly available \cite{Shah:2025zenodo}

\paragraph*{\bf Acknowledgements.} J.N. and M.H. thank Abhishek Kodumagulla for collaboration in the earliest stages of this work. 
V.V. acknowledges fruitful discussions with Volker Koch.
This material is based upon work supported by the National Science Foundation under grants No. PHY-2208724,
and PHY-2116686, and within the framework
of the MUSES collaboration, under grant number No. OAC-
2103680. This material is also based upon work supported
by the U.S. Department of Energy, Office of Science, Office of Nuclear Physics, under Award Numbers DE-SC0022023 and DE-SC0026065, and by the National Aeronautics and Space Agency (NASA)  under Award Number 80NSSC24K0767. 
M.H. was supported in part by  Universidade Estadual do Rio de Janeiro, within the Programa de Apoio à Docência (PAPD).  
J.N. is partly supported by the U.S. Department of Energy, Office of Science, Office for Nuclear Physics under Award No. DE-SC0023861. 
\bibliography{main-v1}
\begin{appendix}
\widetext 
\setcounter{equation}{0}
\setcounter{figure}{0}
\renewcommand{\theequation}{A.\arabic{equation}}
\renewcommand{\thefigure}{A.\arabic{figure}}
\makeatletter
\section*{SUPPLEMENTAL MATERIAL}
\subsection*{Expansion coefficients}

The coefficients $\coef_{2n}$ of the constant entropy density expansion 
\begin{equation}
    T_s(\mu_B; T_0) \approx T_0 + \sum_{n=1}^N \coef_{2n}(T_0) \,\frac{\mu_B^{2n}}{(2\,n)!} + \mathcal{O}\left(\mu_B^{2(N+1)}\right),
    \label{eq:expSsup}
\end{equation}
correspond to the derivatives of the temperature $T$ with respect to $\mu_B$ while holding the entropy density $s$ fixed,
\begin{equation}
\coef_{2n}(T_0) = \left. \left(\frac{\partial^{2n} T}{\partial \mu_B^{2n}}\right)_s \right|_{T = T_0, \mu_B = 0}.
\end{equation}
To evaluate these derivatives, we write the entropy density differential $ds$
\begin{equation}
ds = (\partial s / \partial T)_{\mu_B} dT + (\partial s / \partial \mu_B)_{T} d \mu_B,
\end{equation}
or
\begin{equation}
dT = \frac{1}{(\partial s / \partial T)_{\mu_B} } ds - \frac{(\partial s / \partial \mu_B)_{T}}{(\partial s / \partial T)_{\mu_B}} d \mu_B,
\end{equation}
thus
\begin{align}
\label{eq:s1}
\left(\frac{\partial T}{\partial \mu_B}\right)_s & = -\frac{(\partial s / \partial \mu_B)_{T}}{(\partial s / \partial T)_{\mu_B}} 
= -\frac{(\partial \rho_B / \partial T)_{\mu_B}}{(\partial s / \partial T)_{\mu_B}} = -\frac{\rho_B'}{s'}.
\end{align}
This expression determines the change of the temperature with respect to $
\mu_B$ along a contour of constant entropy density.
Given that $\rho_B = 0$ for all temperatures at $\mu_B = 0$, 
this derivative vanishes at $\mu_B = 0$:
\begin{equation}
\alpha_1(T_0) =  \left. \left(\frac{\partial T}{\partial \mu_B}\right)_s \right|_{T=T_0,\mu_B = 0} = 0.
\end{equation}
The coefficient $\coef_2$ reads
\begin{equation}
\coef_2(T_0) = \left. \left(\frac{\partial^2 T}{\partial \mu_B^2}\right)_s \right|_{T=T_0,\mu_B = 0}.
\end{equation}
We apply a derivative $(\partial / \partial \mu_B)_s$ to Eq.~\eqref{eq:s1} and evaluate it using the chain rule, yielding 
\begin{align}
\left(\frac{\partial^2 T}{\partial \mu_B^2}\right)_s & =
\left[ \frac{\partial (\partial T / \partial \mu_B)_s}{\partial \mu_B} \right]_{T} + \left[ \frac{\partial (\partial T / \partial \mu_B)_s}{\partial T} \right]_{\mu_B} \left(\frac{\partial T}{\partial \mu_B}\right)_s
\nonumber \\
& = -\frac{2T \chi_2^{B} + T^2 \chi_2^{B'}}{s'} + \frac{\rho_B' \rho_B''}{[s']^2} + \left[ \frac{\rho_B' s''}{(s')^2} - \frac{\rho_B''}{s'} \right] \left(\frac{\partial T}{\partial \mu_B}\right)_s,
\label{eq:app:C2_chainrule}
\end{align}
where we took into account that $\rho_B = T^3 \chi_1^B$ and $(\partial \chi_1^B / \partial \mu_B)_T = \chi_2^B / T$.
At $\mu_B = 0$, only the first term is non-zero, giving
\begin{equation}
\label{eq:app:C2}
\coef_2(T_0) = \left. \left(\frac{\partial^2 T}{\partial \mu_B^2}\right)_s \right|_{T=T_0,\mu_B = 0} = -\frac{2T_0 \chi_2^{B}(T_0) + T_0^2 \chi_2^{B'}(T_0)}{s'(T_0)}.
\end{equation}

\subsection*{\bf Lattice QCD input and its description}

\begin{figure*}
\centering
    \includegraphics[width=0.4\textwidth]{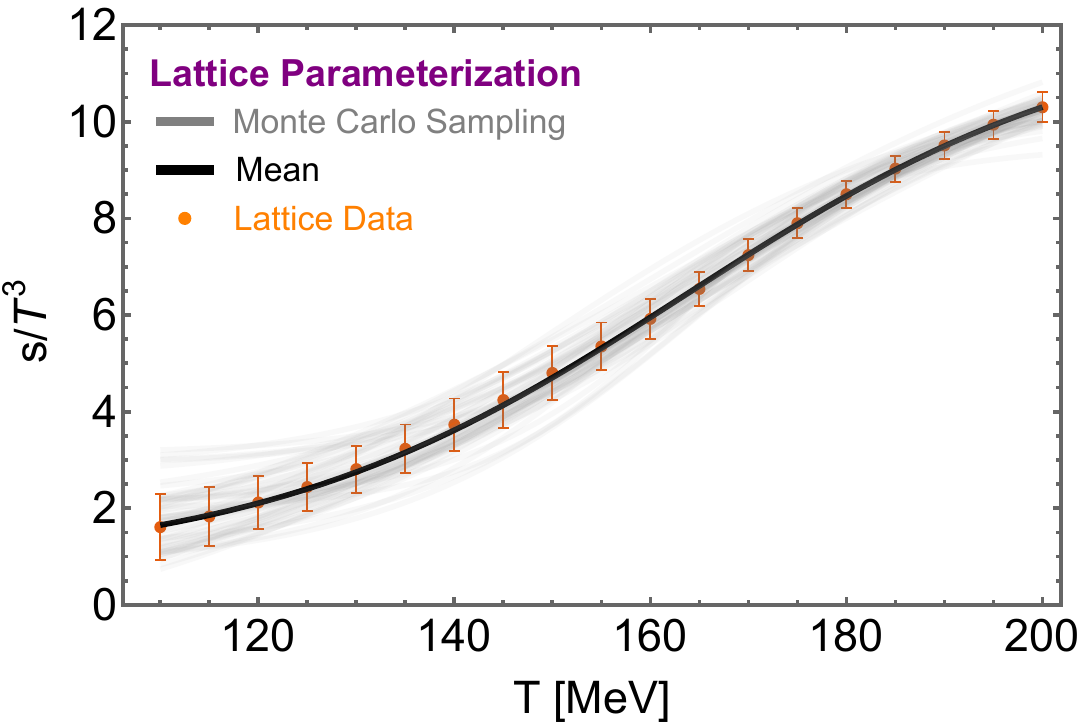} 
    \hspace{1cm} 
    \includegraphics[width=0.4\textwidth]{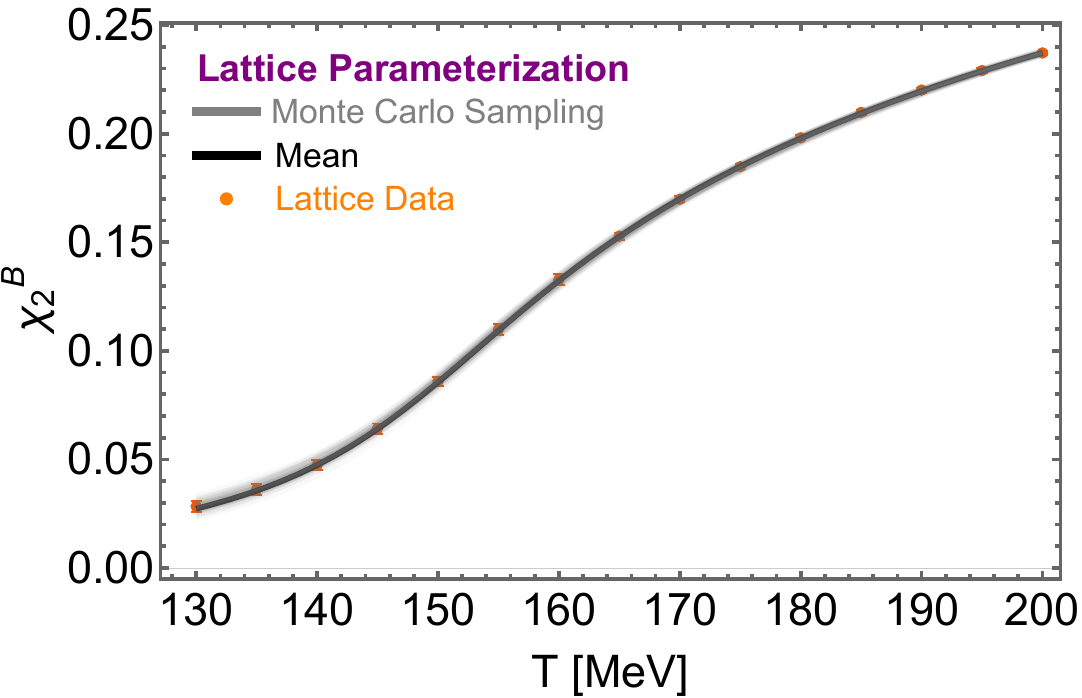}
    \caption{\justifying Scaled entropy density $s/T^3$ (left panel) and second order baryon number susceptibility $\chi_2^B$ (right panel) as functions of the temperature at $\mu_B = 0$. 
    The orange symbols with error bars depict the lattice QCD data of the Wuppertal-Budapest collaboration~\cite{Borsanyi:2013bia,Borsanyi:2021sxv}.
    The solid black line corresponds to the parameterization of the lattice data using the mean parameter values, while the hazed black lines correspond to different Monte Carlo samples of the parameters reflecting the uncertainties in the lattice data.
    }
    \label{fig:latticedata}
\end{figure*}

Our method requires entropy density $s(T)$ and baryon number susceptibility $\chi_2^B(T)$ at vanishing chemical potential $\mu_B = 0$ as input.
We utilize continuum-extrapolated lattice data for $s(T)$ and $\chi_2^B(T)$ of the Wuppertal-Budapest collaboration from Refs.~\cite{Borsanyi:2013bia} and~\cite{Borsanyi:2021sxv}, respectively. These lattice QCD data are shown in Fig.~\ref{fig:latticedata}.

\paragraph*{\bf Entropy density.}

As our method requires the use of temperature derivatives, we parametrize the lattice QCD input and its uncertainties.
For the entropy density, we use the following function
\begin{equation}
\label{eq:sT3param}
    \frac{s}{T^3} = a \cdot \tanh \left( \frac{T - T_0}{d} \right) + b.
\end{equation}
To determine the values of the parameters and their uncertainties, we perform a fit of the lattice data at temperatures $110 < T < 200$~MeV through $\chi^2$ minimization.
The expression for $\chi^2_s$ (the index $s$ indicates that it is the $\chi^2$ for the entropy) reads
\begin{equation}
\chi^2_s = \sum_{i,j=1}^N [(s/T^3)_i^{\rm param}-(s/T^3)_i^{\rm lattice}] (\Sigma^{-1}_{s})_{ij}^{\rm lattice}[(s/T^3)_j^{\rm param}-(s/T^3)_j^{\rm lattice}].
\end{equation}
Here the sums runs over all lattice data points, where $i = 1$ corresponds to the lowest temperature, $T_1 = 110$~MeV and $i = N$ to the highest temperature, $T_N = 200$~MeV, used in the fit. 
All temperature points are equally spaced, with a step of $\Delta T = 5$~MeV.
$\Sigma_s$ is the covariance matrix describing the lattice errors and correlations among them.
It is taken in the form
\begin{equation}
(\Sigma_s)_{ij}^{\rm lattice} = (\sigma_s)_i (\sigma_s)_j\, \Gamma_s^{|i-j|}.
\end{equation}
Here $0 < \Gamma_s < 1$ is the covariance coefficient correlating the uncertainties between lattice data points at different temperatures.
This choice was considered in Ref.~\cite{Hippert:2023bel} and implies stronger correlations between data points that are closer in temperatures.
It was also indicated in Ref.~\cite{Hippert:2023bel} that a value of $\Gamma_s = 0.84$ describes the lattice data the best, and we use this value in the present analysis.

\begin{table}[ht]
\centering
    \begin{tabular}{|c|c|}
    \hline
        Parameter & Value \\
        \hline
        a & 5.65608 \\ 
        b & 6.43026 \\ 
        $T_0$ & 163.681 \\
        d & 43.3516 \\
        \hline
    \end{tabular}
    \hspace{2cm}%
    \begin{tabular}{|c|c|c|c|c|}
    \hline
      Covariance   & a & b & $T_0$ & d \\
    \hline
      a   & 1.24081 & -0.133316 & -1.5958 & 13.4657 \\
         \hline
      b   & -0.133316 & 0.232124 & 2.08816 & -0.507437 \\
         \hline
      $T_0$  & -1.5958 & 2.08816 & 24.4358 & -14.2532\\
         \hline
      d   & 13.4657 & -0.507437 & -14.2532 & 161.338 \\
         \hline
    \end{tabular}
\caption{\justifying
Mean values of parameters describing the lattice QCD data on $s/T^3$ via parametrization~\eqref{eq:sT3param}~(left table) and their covariance matrix~(right table). The units for $T_0$ are MeV.
}
\label{tab:param:sT3}
\end{table}

Table~\ref{tab:param:sT3} depicts the mean parameter values and the resulting covariance matrix obtained from fitting the lattice QCD data through $\chi^2_s$ minimization.
The left panel of Fig.~\ref{fig:latticedata} shows the comparison of the resulting parametrization with lattice QCD data.
The solid black line corresponds to the mean parameter values, showing excellent agreement with the mean values of the lattice data.
The 50 hazy gray lines depict the results of the parametrization by sampling the parameter sets from a multi-Gaussian distribution with the mean vector and covariance matrix listed in Table~\ref{tab:param:sT3}.
The resulting spread of the lines accurately reflects the uncertainties in the lattice data.

\paragraph*{\bf Baryon number susceptibility.}
We parametrize the baryon number susceptibility $\chi_2^B(T)$ at $\mu_B = 0$ in the following form
\begin{equation}
\chi_2^B(T) = d_0 \left( \frac{2 \cdot m_p}{\pi \cdot x} \right)^{\frac{3}{2}} \cdot \frac{\exp\left(-\frac{m_p}{x}\right)}{1 + \left(\frac{x}{d_1}\right)^{d_2}} + d_3 \cdot \frac{\exp\left(-\frac{d_5^4}{x^4}\right)}{1 + \left(\frac{x}{d_1}\right)^{-d_2}}.
\label{eq:chi2_para}
\end{equation}
Here $x = T/(200~\text{MeV})$ and $m_p = (938/200) \approx 4.7$ is the mass of the proton mass in units of 200~MeV.
This is a modified version of the parametrization used in Ref.~\cite{Kahangirwe:2024cny}, where we introduced an additional parameter $d_0$ and removed the parameter $d_4$ by setting it to zero.
This modification leads to a more accurate description of the lattice data~\cite{Borsanyi:2021sxv} on $\chi_2^B$ and its temperature derivative at $T \lesssim 145$~MeV.
As in the case of entropy density, the determination of parameters and their covariance matrix proceeds by minimizing $\chi^2_{\chi_2}$:
\begin{equation}
\chi^2_{\chi_2} = \sum_{i,j=1}^N [(\chi_2^B)_i^{\rm param}-(\chi_2^B)_i^{\rm lattice}] (\Sigma^{-1}_{\chi_2})_{ij}^{\rm lattice}[(\chi_2^B)_j^{\rm param}-(\chi_2^B)_j^{\rm lattice}],
\end{equation}
where the covariance matrix of the lattice uncertainties reads
\begin{equation}
(\Sigma_{\chi_2})_{ij}^{\rm lattice} = (\sigma_{\chi_2})_i (\sigma_{\chi_2})_j\, \Gamma_{\chi_2}^{|i-j|},
\end{equation}
with the correlation coefficient $\Gamma_{\chi_2} = 0.84$~\cite{Hippert:2023bel}.
In contrast to the entropy density data, the lattice data for $\chi_2^B$ start at $T_1 = 130$~MeV in steps of $\Delta T = 5$~MeV.
We use the lattice data in a range $130 \leq T \leq 200$~MeV in the fitting procedure.
The resulting mean values of the parameters and their covariance matrix are listed in Table~\ref{tab:param:chi2}.

\begin{table}[ht]
\centering
    \begin{tabular}{|c|c|}
    \hline
        Parameter & Value \\
        \hline
        $d_0$ & 3.60763 \\ 
        $d_1$ & 0.750215 \\ 
        $d_2$ & 21.1553 \\
        $d_3$ & 0.330518 \\
        $d_5$ & 0.758584 \\
        \hline
    \end{tabular}
    \hspace{1cm}%
    \begin{tabular}{|c|c|c|c|c|c|}
    \hline
      Covariance   & $d_0$ & $d_1$ & $d_2$ & $d_3$ & $d_5$\\
    \hline
      $d_0$   & 0.136045 & 0.00112188 & 0.630763 & 0.000696749 & 0.00126876 \\
         \hline
      $d_1$   & 0.00112188 & 0.0000375329 & 0.00251407 & $-6.53388 \cdot 10^{-6}$ & -0.0000106728
 \\
         \hline
      $d_2$  & 0.630763 & 0.00251407 & 7.31758 & 0.00923282 & 0.0167017
 \\
         \hline
      $d_3$   & 0.000696749 & $-6.53388 \cdot 10^{-6}$ & 0.00923282 & 0.0000223041 & 0.0000379904
 \\
         \hline
      $d_5$   & 0.00126876 & -0.0000106728 & 0.0167017 & 0.0000379904 & 0.0000664791
 \\
         \hline
    \end{tabular}
\caption{\justifying
Mean values of parameters describing the lattice QCD data on $\chi_2^B$ via parametrization~\eqref{eq:chi2_para}~(left table) and their covariance matrix~(right table). 
}
\label{tab:param:chi2}
\end{table}

The right panel of Fig.~\ref{fig:latticedata} shows the comparison of our parametrization with the lattice data.
As in the case of $s/T^3$, the comparison indicates that our parametrization captures the behavior of both the means and the lattice errors accurately.

\paragraph*{\bf Expansion coefficient $\coef_2$.}

\begin{figure*}
\centering
    \includegraphics[width=0.32\textwidth]{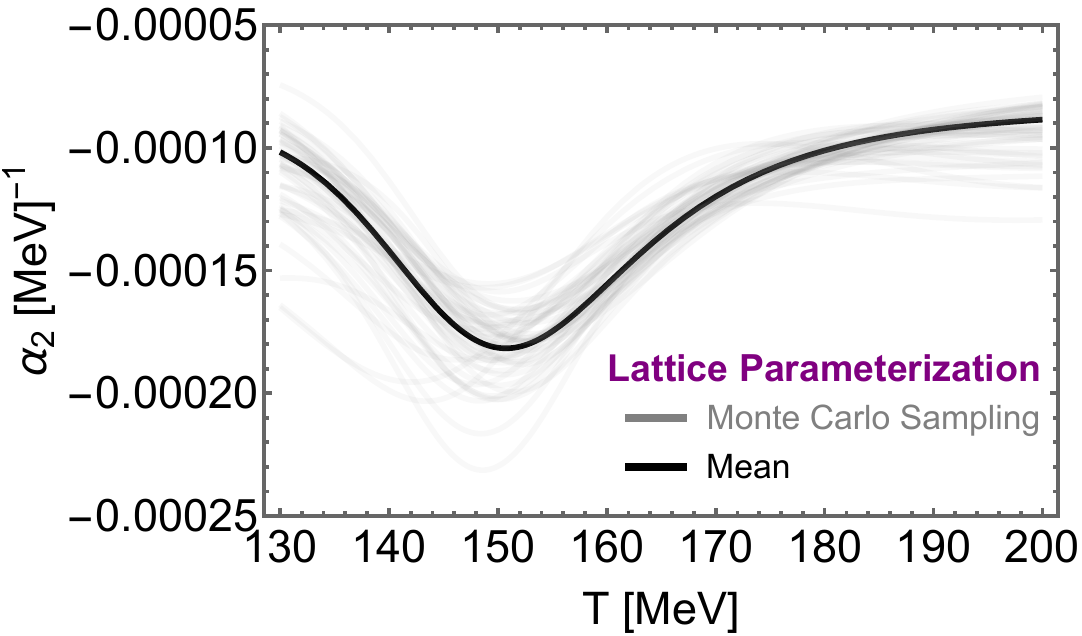}
    \includegraphics[width=0.32\textwidth]{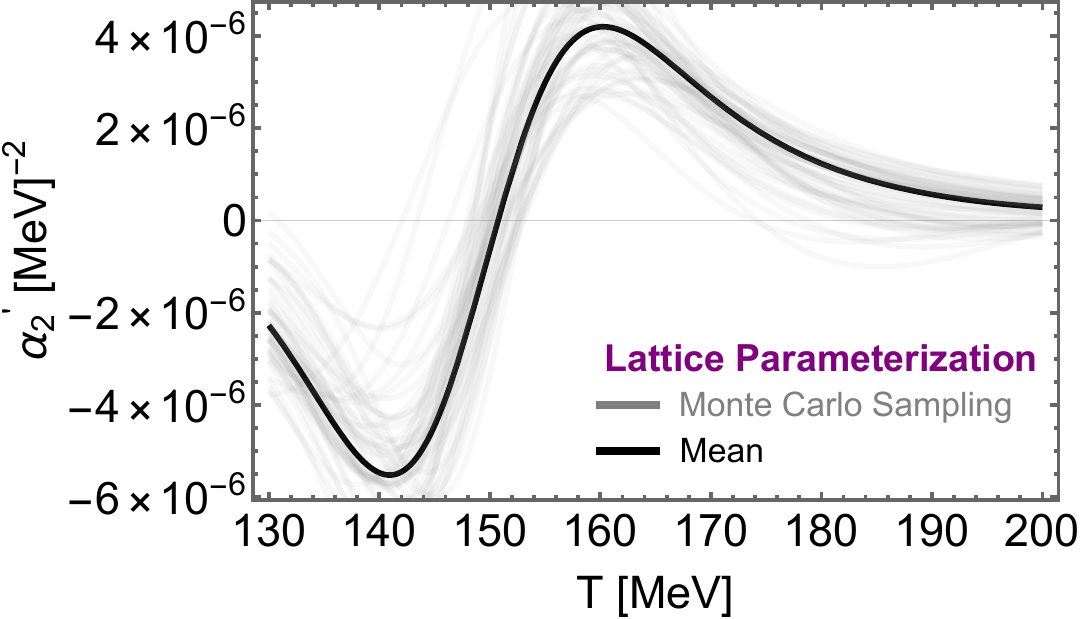}
    \includegraphics[width=0.32\textwidth]{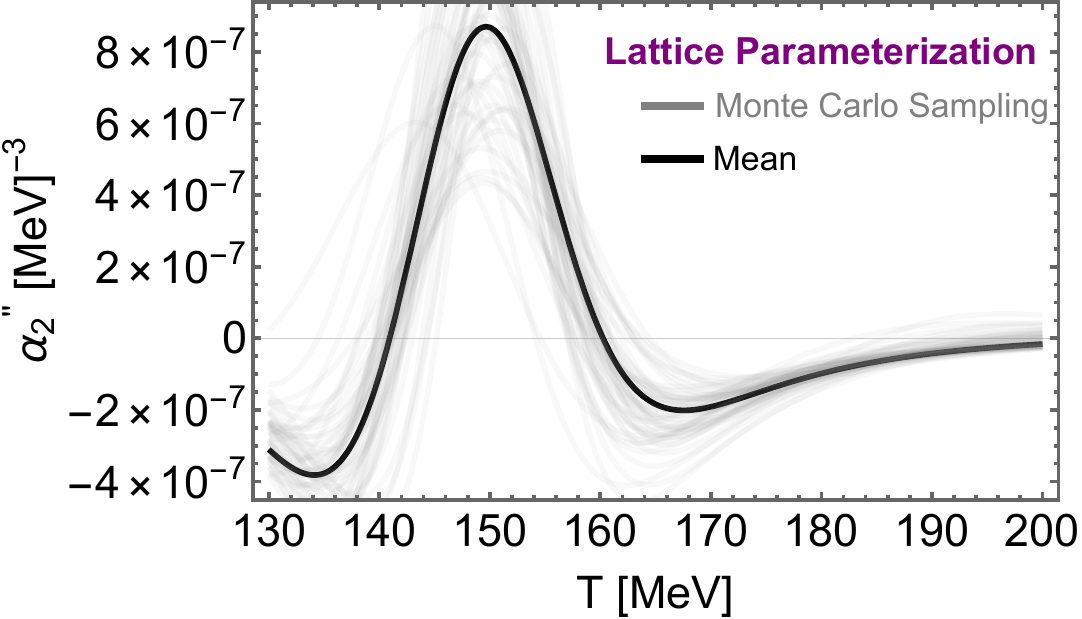}
    \caption{\justifying 
    Temperature dependence of the second order expansion coefficient $\coef_{2}$~(left panel), and its first~(middle panel) and second~(right panel) temperature derivatives obtained using the parametrization of the lattice data.
    The black lines correspond to mean parameter values while hazy gray lines correspond to the Monte Carlo sampling of parameter values.    
    }
    \label{fig:C2}
\end{figure*}

Figure~\ref{fig:C2} shows the temperature dependence of the expansion coefficient $\coef_2$ and its first and second derivatives, as obtained through Eq.~\eqref{eq:app:C2} and using parameterizations~\eqref{eq:sT3param} and~\eqref{eq:chi2_para} for $s/T^3$ and $\chi_2^B$, respectively.
As in Fig.~\ref{fig:latticedata}, the solid black curve corresponds to the mean parameter values, while the hazy gray curves correspond to Monte Carlo sampling of the parameters and reflect the propagation of the lattice data uncertainties.

\subsection*{Coexistence line}

The coexistence curve $T_{cc}(\mu_B)$ at $\mu_B > \mu_{B,c}$ can be determined by considering the Gibbs criteria for phase equilibrium, that state that the two phases are in equilibrium if the chemical potentials, temperature, and pressures~(or, equivalently, the Gibbs free energy) are equal
$$
\mu_{B,1} = \mu_{B,2}, \qquad T_1 = T_2, \qquad P_1 = P_2.
$$
Let us consider a line of constant $\mu_B > \mu_{B,c}$, where points 1 and 2 correspond to the two phases.
Given that $dP = s dT$ at a constant $\mu_B$, one can write the change in pressure along the $\mu_B = \rm const.$ line between points 1 and 2 as
\begin{align}
\Delta P_{12} = \int_1^2  s(T) dT.
\end{align}
Using the integration by parts, one can rewrite this as
\begin{align}
\Delta P_{12} = s_2 T_2 - s_1 T_1 - \int_{s_1}^{s_2} T(s) ds.
\end{align}
Due to phase equilibrium condition, $\Delta P_{12} = P_2 - P_1 = 0$ and $T_1 = T_2 = T_{\rm cc} (\mu_B)$.
One also has $s_{1,2} = s_{1,2}(T_{\rm cc})$ and $T(s) = T_s[\mu_B;T_0(s)]$.
Phase coexistence criteria therefore reduce a Maxwell construction of equal areas for entropy density
\begin{align}
    \int_{s_1(T_{cc})}^{s_2(T_{cc})} \{T_s[\mu_B;T_0(s)] - T_{cc}(\mu_B)\} ds = 0.
\end{align}


\subsection*{Smoothing splines}

One might question how our results depend on the particular parametrization we choose for the entropy and the second baryon susceptibility. 
To address this concern, we have also carried out a more agnostic analysis, using splines to interpolate the lattice results and Monte Carlo to propagate uncertainties. 
This analysis is discussed below.

For the error propagation, we create random pseudo-data such that its average reproduces the lattice results and its statistical errors reproduce the corresponding uncertainties.  
For each value of the temperature, $T_i$ , with $i=1,\ldots,N$, we take 
\begin{equation}
    q_i = ({q})_i^{\textrm{lattice}} + (\xi_q)_i
\end{equation}
where $q=s/T^3,\,\chi_2^B$. 
The statistical noise $(\xi_q)_i$ is randomly sampled from a Gaussian, of average $\langle 
(\xi_q)_i \rangle = 0$ and covariance 
\begin{equation}
    \langle (\xi_q)_i \, (\xi_{q'})_j\rangle = \delta_{q q'}\, (\sigma_q)_i (\sigma_{q'})_j\, \Gamma_q^{|i-j|}\,,
\end{equation}
where $(\sigma_q)_i$ are the lattice QCD uncertainties and we take the correlation coefficient $\Gamma_q =0.84$ \cite{Hippert:2023bel}. 
For each quantity $q=s/T^3,\,\chi_2^B$, we sample a thousand pseudo-data realizations $\{q_i\}$.
\begin{figure*}
\centering
    \includegraphics[width=0.45\textwidth]{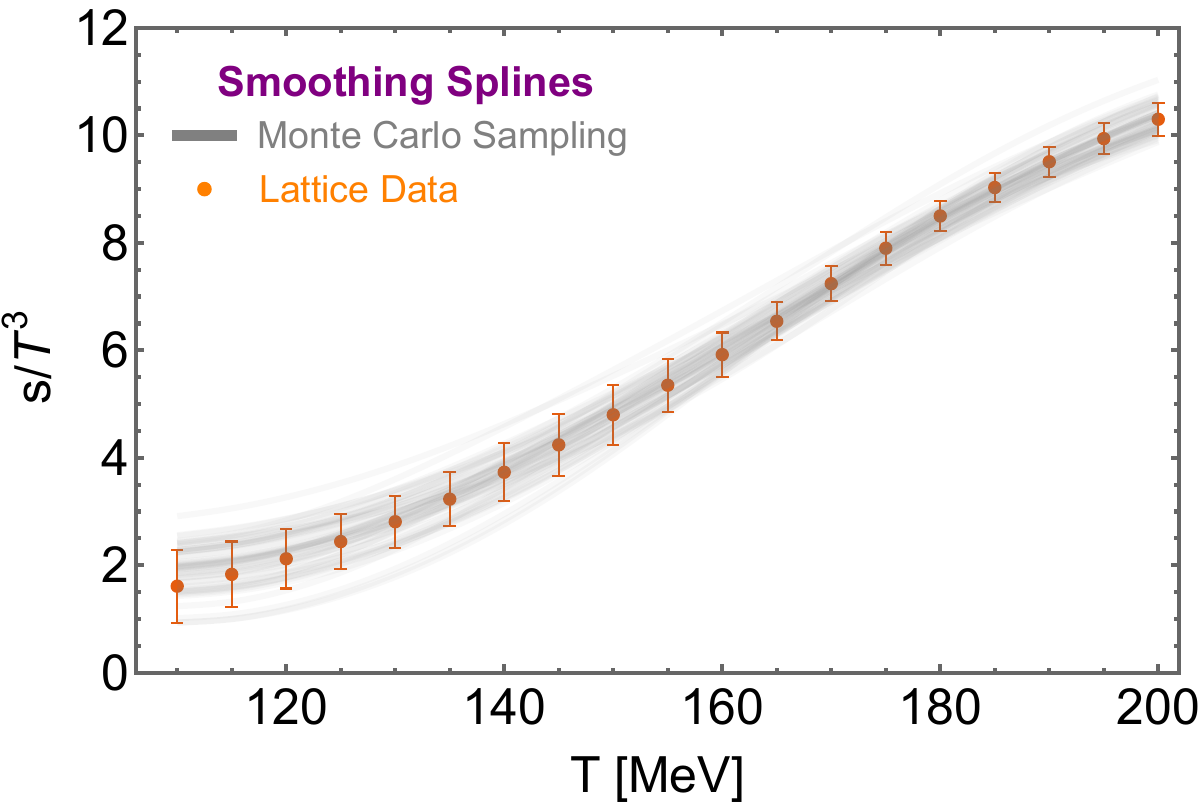}
    \includegraphics[width=0.45\textwidth]{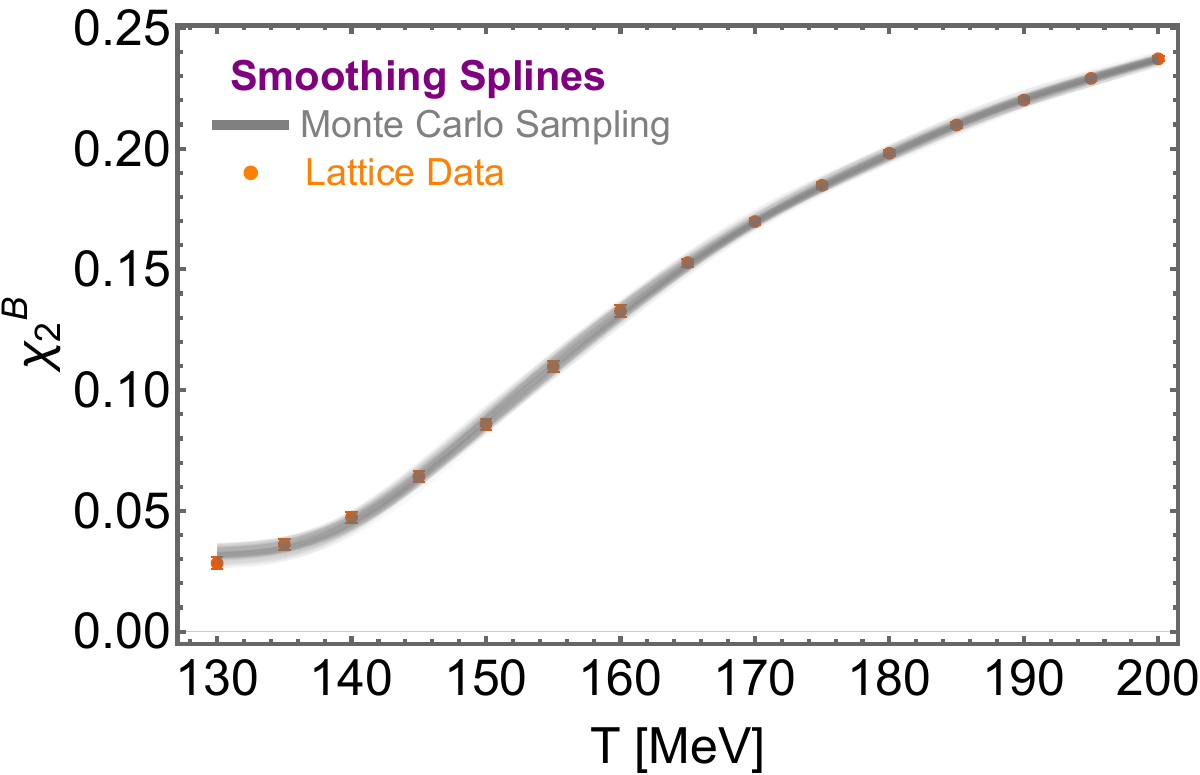}
    \caption{\justifying 
    Same as Fig.~\ref{fig:latticedata} but using smoothing splines for describing the lattice data instead of parametrization.
    } 
    \label{fig:latticedata_splines}
\end{figure*}
We then model the random pseudo-data sample with splines, so we can perform differentiation in a well-controlled way. 
A naive interpolation with splines would go through each point in the random pseudo-data, leading to artifacts in the derivatives due to  overfitting. 
Thus, we instead fit the pseudo-data with cubic smoothing splines
\cite{de1978practical}. 
We take natural cubic splines $f_q(T)$, with knots at the different $T_i$, constructed to minimize \cite{de1978practical}
\begin{equation}
    \tilde\chi_q[f_q]^2 = p_q \,\sum_{i=1}^N \left(\frac{q_i-f_q(T_i)}{(\sigma_q)_i}\right)^2
    + (1-p_q) \int_{T_1}^{T_N} dT \,(f_q''(T))^2\,,
    \label{eq:smoothingcost}
\end{equation}
which favors smoother splines. 
Following \cite{de1978practical}, we pick the parameter $p_q$ in Eq.~\eqref{eq:smoothingcost}  just large enough that 
\begin{equation}
    S[f_q]\equiv \sum_{i=1}^N \left(\frac{q_i-f_q(T_i)}{(\sigma_q)_i}\right)^2 \leq N\,.
\end{equation}
This procedure leads to an ensemble of functions that reproduce, to good accuracy, the lattice QCD results and uncertainties at $T=T_i$, $i=1,\ldots,N$, whilst displaying smooth, well-behaved derivatives. The smooth splines for entropy density and baryon number susceptibility are shown in Fig. \ref{fig:latticedata_splines}, where the smoothed splines performed on the pseudo data samples are in hazy gray lines compared to the lattice data.
One can see good agreement with the lattice data and their errors, with the possible exception of the lowest temperature point for $\chi_2^B$, which may be reflecting splining artefacts at the boundary.

\begin{figure*}
\centering
    \includegraphics[width=0.32\textwidth]{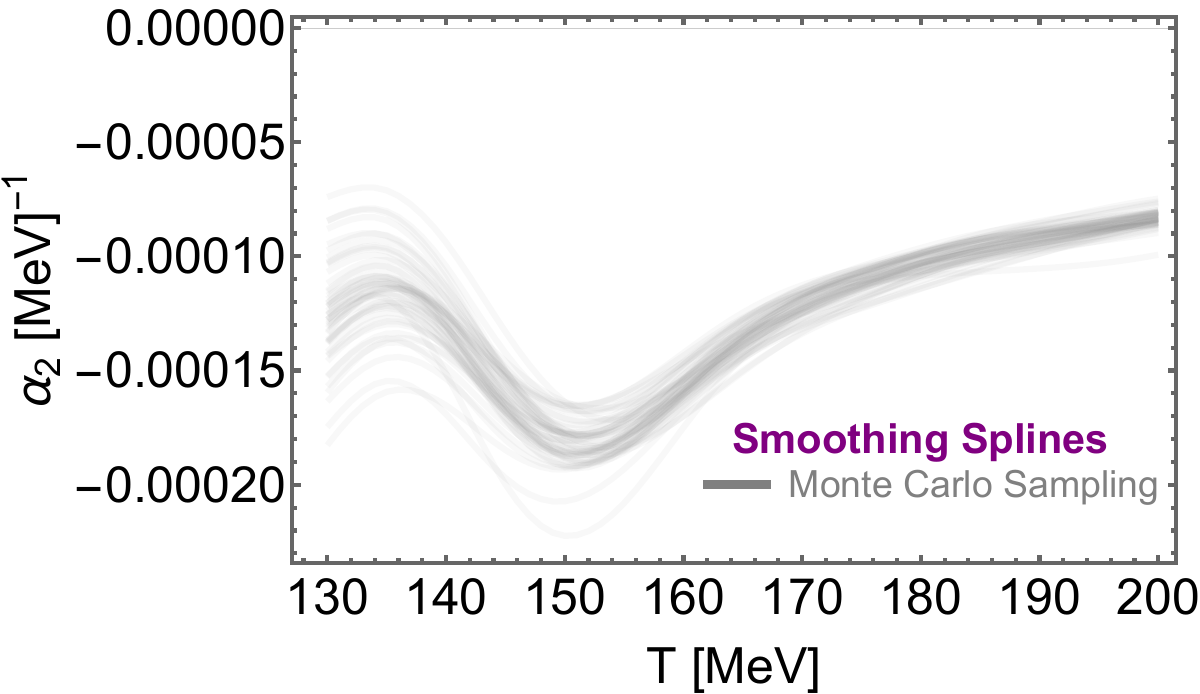}
    \includegraphics[width=0.32\textwidth]{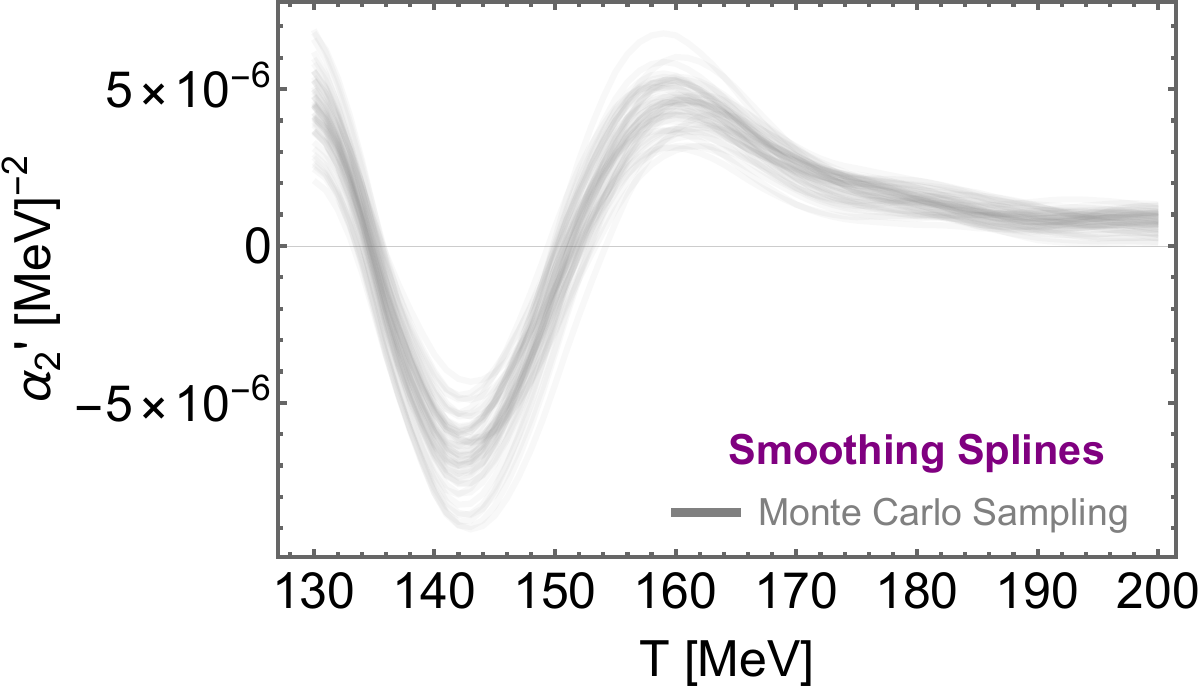}
    \includegraphics[width=0.32\textwidth]{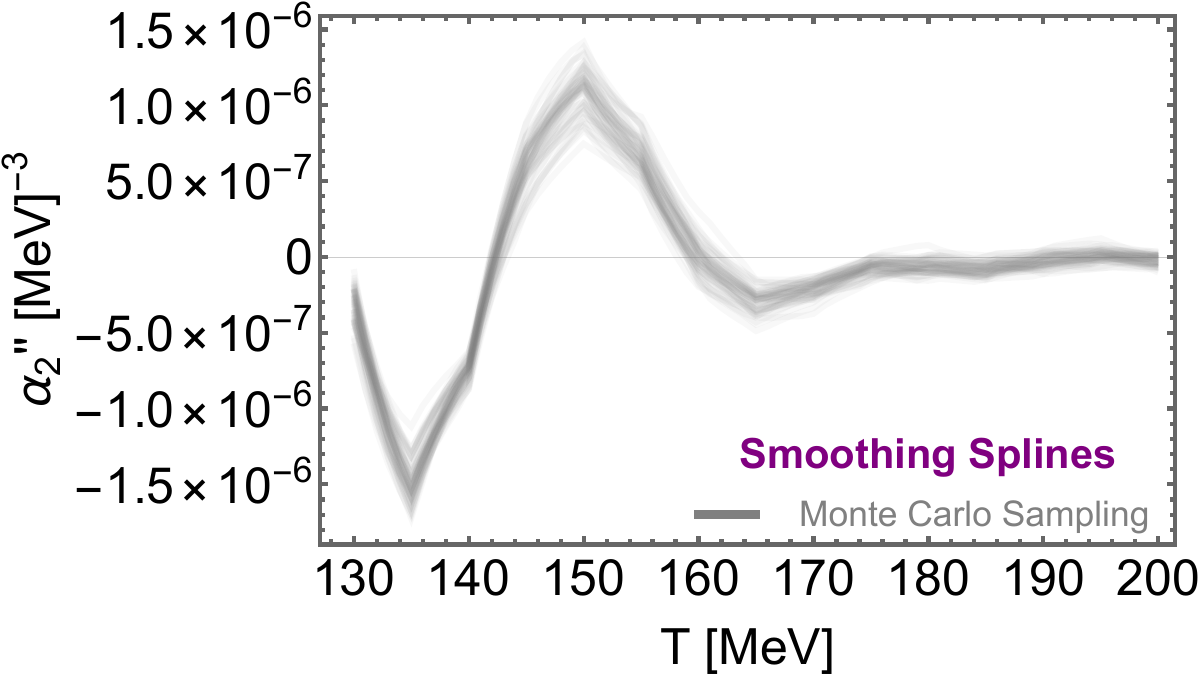}
    \caption{\justifying 
    Same as Fig.~\ref{fig:C2} but using smoothing splines for describing the lattice data instead of parametrization.
    }
    \label{fig:C2_splines}
\end{figure*}

\begin{figure}[!h]
    \centering
    \includegraphics[width=0.45\linewidth]{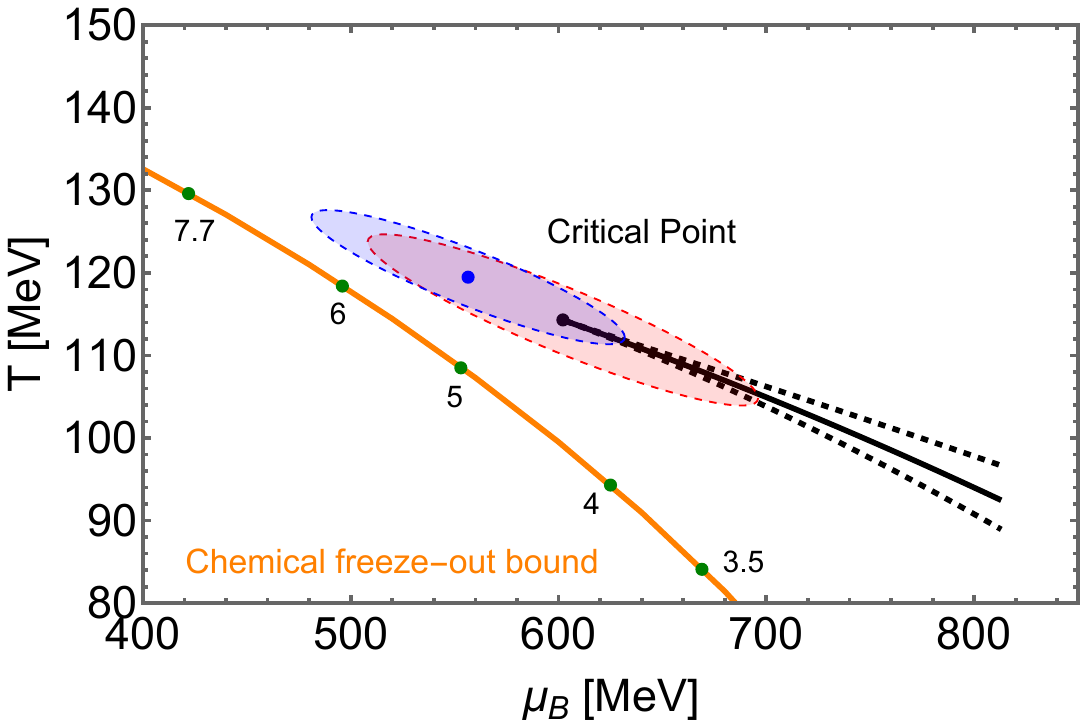}
    \caption{\justifying \small
    Same as Fig.~\ref{fig:CP} of the main text, but additionally showing the 68\% confidence ellipse~(blue ellipse) for the CP location obtained using smoothing splines. The green points show various chemical freeze-out points for different collision energies. 
    }
    \label{fig:CP_splines}
\end{figure}

From the splines obtained for each of the pseudo-data sets, we compute the coefficient $\coef_2(T_0)$ and its temperature derivatives given in Fig. \ref{fig:C2_splines}.
These show good agreement with the results obtained using the parametrization, again, with the possible exception of the lowest temperatures $T \lesssim 135$~MeV at the boundary.
We use $\coef_{2}$ and its derivatives
to determine the location of the CP, computing the mean and standard deviation of $T_c$ and $\mu_{B,c}$ to obtain 
\begin{equation}
    (T_c, \mu_{B,c}) = (119.5 \pm 5.4,\,556.5\pm 49.8) \text{\MeV},
\end{equation}
which agrees with our main result within one sigma. This can be seen in Fig. \ref{fig:CP_splines}, where the blue ellipse is the covariance ellipse at 68\% confidence level and reflects the lattice QCD input's error propagation. To calculate this ellipse, a thousand samples were generated and splining was performed based on Eq. \eqref{eq:smoothingcost}. The covariance ellipse is above the lowest temperature bound of the chemical freezeout curve described by the HRG input given in \cite{Lysenko:2024hqp}, while also being in agreement with the covariance ellipse (red) obtained using the parametrization shown in Fig. \ref{fig:C2}.

\end{appendix}
\end{document}